\documentclass[fleqn,twoside]{article}
\usepackage{espcrc2}

\usepackage{graphicx}
\usepackage[figuresright]{rotating}

\newcommand{\AmS}{{\protect\the\textfont2
  A\kern-.1667em\lower.5ex\hbox{M}\kern-.125emS}}

\title{QCD meson spectrum in the large $N_C$ limit}

\author{G. Grignani\address[MCSD]{Dipartimento di Fisica e Sezione I.N.F.N. Universit\`a di Perugia, \\
        Via A. Pascoli, 06123 Perugia, Italy},%
        D. Marmottini\addressmark
and
        P. Sodano\addressmark}

\begin{document}

\begin{abstract}
The  low energy  mass spectrum of QCD in the large $N_C$ limit is
computed.  The low-lying states in  the meson spectrum are
explicitly evaluated up to the fourth order  in the strong
coupling perturbative expansion. The 't Hooft limit is smooth and
the meson masses  are in  very good agreement with the
experimental values. \vspace{1pc}
\end{abstract}

\maketitle

\section{INTRODUCTION}

Since  the seminal  work of  't Hooft
\cite{'tHooft:1973jz,'tHooft:1974hx} the  large $N_C$ limit has
played an increasingly important role in studying gauge theories
in the continuum, on the lattice (see for
example~\cite{Teper:2002kh} and references therein) and the
duality between gauge and string theory \cite{Aharony:1999ti}.
However, there are still no estimates of physically observable
quantities, where the large $N_C$ limit can be tested against
experiments. Here we present the results of an explicit and
successful calculation of the meson masses, in the large $N_C$
limit~\cite{dona}. In~\cite{dona} we extend to an arbitrary number
of colors the procedure used by Banks et al.~\cite{Banks:1976ia}
in their, by now famous, evaluation of the meson spectrum and then
take the large $N_C$ limit. The 't Hooft limit is smooth
\cite{Witten:1979kh} and all the divergent terms in the
thermodynamic limit cancel against the ground state vacuum energy.

The theory is studied in the  strong coupling limit  using
staggered fermions  and  the Hamiltonian  approach to lattice
gauge theory. We  use $x=1/{g^2}$ as the expansion parameter and
the continuum limit is  extrapolated using  Pad\'e approximants.
The  computation is performed by taking as the  ground state that
of the antiferromagnetic Ising  model which is  gauge invariant
and breaks  chiral symmetry. The gauge invariant eigenstates of
the unperturbed Hamiltonian are  used  as a basis for the
perturbative expansion. A similar approach has been successfully
used to study the meson spectrum of some simpler toy models like
the Schwinger models \cite{Berruto:1997jv} and the two dimensional
't Hooft model~\cite{Berruto:1999ir}.

\section{HAMILTONIAN FORMULATION}

We consider the Hamiltonian formulation  of lattice gauge theory
where the time is
continuous  and the space is discretized on a 3-dimensional cubic
lattice  with  sites $x=(x_1,x_2,x_3)$,  where  $x_i$ are  integers.
The Hamiltonian reads
\begin{eqnarray}
\nonumber
&& H=\frac{1}{2a}\left(g^2\sum_{[x,n]}E^a[x,n]^2+\frac{1}{g^2}\sum_{[x,i,j]}Tr(UUUU)\right.\\
&&
\left.+\sum_{[x,n]}\eta(\hat{n})\Psi^{\dagger}_{A}(x+\hat{n})U_{AB}[x,n]\Psi_B(x)+h.c.\right)
\end{eqnarray}
($TrUUUU=Tr(\prod_{plaq[x,i,j]}U_{[x,i]}$)) where
$$ \eta(\hat{x})=(-1)^{z} \quad \eta(\hat{y})=(-1)^{x} \quad \eta(\hat{z})=(-1)^{y}$$
are the Dirac matrices for staggered
fermions~\cite{Kluberg-Stern:1983dg} and $a$ is the lattice
spacing. The gauge field $U[x,n]$ is associated with  the link
$[x,n]$ and it is  a group element in the fundamental
representation  of $SU(N_C)$. The electric field operator
$E^a[x,n]$ is defined  on the link and it obeys the algebra
$$[E^a[x,n],E^b[y,j]]=if^{abc}E^c[x,n]\delta([x,n]-[y,j])$$
and $E[x,n]=E^a[x,n]T^a$, with $T^a$, $a=1,...,N_C^2-1$ the
generators of the Lie algebra of $U(N_C)$. It generates the
left-action of the Lie algebra on $U[x,n]$
$$[E^a[x,n],U[y,j]]=-T^aU[x,n]\delta([x,n]-[y,j])$$
The fermion fields $\Psi$ are defined on the lattice sites and obey the relation
$$\{\Psi_A(x),\Psi_B^\dagger(y)\}=\delta_{AB}\delta(x-y)$$
Each term in the Hamiltonian is gauge invariant under static gauge
transformation. The Hamiltonian is also invariant under some
discrete symmetries which are used  to construct  the operators
generating  the mesons  from the vacuum~\cite{Susskind:1976jm}.
They are:  lattice translation  by a single link, shift along a
face diagonal, parity, G-parity and Charge conjugation.   In
particular,  the invariance  under translation  by a single link
plays the  role of a  discrete chiral invariance of  the theory.
This symmetry is broken by an explicit mass term in the
Hamiltonian.

\subsection{The strong-coupling expansion}

In the strong-coupling expansion $H_0$ is treated as unperturbed
Hamiltonian, while $H_1$ and $H_2$ are the perturbations. $H_1$ is
the interaction Hamiltonian between the quarks and the gauge
fields and $H_2$ is the magnetic Hamiltonian:
\begin{eqnarray}
\nonumber
&&H_0=\frac{g^2}{2a}\sum_{[x,n]}E^a[x,n]^2\\
\nonumber
&&H_1=\frac{1}{2a}\sum_{[x,n]}\eta(\hat{n})\Psi^{\dagger}_{A}(x+\hat{n})U_{AB}[x,n]\Psi_B(x)\\
\nonumber
&&\phantom{H_1=}+h.c.\\
\nonumber &&H_2=\frac{1}{2g^2a}\sum_{[x,i,j]}Tr(UUUU)+h.c.
\end{eqnarray}
The vacuum of the theory is a singlet of the electric field
algebra. It is needed then to find the gauge invariant eigenstates
of $H_0$ with the lowest energy $E_0$. The lowest eigenstates of
$H_0$ with $E_0=0$ are singlets of the electric field algebra and
since they are gauge invariant they must be color singlets. These
states must also conserve charge, that means
$\sum_{x}\rho(x)|0>=0$ where $\rho$ is the fermion number operator
$\rho(x)=\Psi^{\dagger}(x)\Psi(x)-N_C/2$. There are $M!/(M/2)!$
degenerate ground states. The degeneracy can be resolved by
diagonalizing the first nontrivial order in perturbation theory,
that is the second
\begin{equation}
E^{(2)}_0=<0|H_2|0>+<0|H_1\frac{\Pi_0}{E_0-H_0}H_1|0>
\end{equation}
where $<,>={\prod_{[x,n]}\int dU[x,n]}(,)$ is the inner product in
the full Hilbert space of the model. This energy has been computed
by constructing an eigenstates of $H_0$ and using it to evaluate
the functions present in the strong coupling expansion. If a state
$|\Psi>$ is a singlet of the electric field algebra
$E^a[x,n]|\Psi>=0$ then $H_0|\Psi>=0$ and
\begin{equation}
H_0U[x,n]|\Psi>=\frac{g^2}{2a}C_2(N_C)U[x,n]|\Psi>
\end{equation}
where $C_2$ is the Casimir operator of $SU(N_C)$. After
integration over the link variable $U$, the vacuum energy becomes
\begin{equation}
E_0^{(2)}=-K<0|\sum_{[x,n]}[\rho(x+\hat{n})+\frac{N_C}{2}][-\rho(x)+\frac{N_C}{2}]
\end{equation}
where $K=1/g^2aC_2 N_C$. The Hamiltonian has the two degenerate
ground states of the antiferromagnetic Ising model; choosing one
of them breaks the chiral symmetry~\cite{Langmann:hd}. The vacuum
energy has been computed up to the fourth order in the
strong-coupling expansion~\cite{dona}.

\section{THE MESON SPECTRUM}

In the strong coupling expansion the lowest-lying states in the
meson spectrum are those consisting of a quark and antiquark at
opposite ends of a single link. For a given meson the wavefunction
may be determined by taking the quark bilinear with the desired
transformation properties, writing it in point-separated lattice
form using the discrete symmetries of the theory and applying it
to the vacuum.  The mesons we have considered are $\pi_0$, $\rho$,
$\omega$, $b_1$, $a_1$, $f_2$ and $f_0$.  All the meson are
degenerate at the lowest order and their energy is
$E_M^{(0)}=g^2C_2/2a$. The meson energy has been computed up to
the fourth order in the perturbative expansion using a procedure
similar to the one used in the evaluation of the vacuum energy. If
one computes the meson energy up to the fourth order, one gets
$$E_M=\frac{g^2C_2}{2a}+E_M^{(2)}+E_M^{(4)}$$
(where $a$ is the lattice spacing). If one rescales the coupling
constant according to the 't Hooft prescription $g^2N_C
\rightarrow g^2$ (large $N_C$ with $g^2N_C$ fixed) one gets for
the meson masses:
\begin{eqnarray}
\nonumber
&& m_{\pi_0}=E_{\pi_0}-E_0=\frac{1}{a\sqrt{y}}[\frac{1}{4}+6y-171y^2]\\
\nonumber
&& m_{\omega}=E_{\omega}-E_0=\frac{1}{a\sqrt{y}}[\frac{1}{4}+6y-171y^2]\\
\nonumber
&& m_{\rho}=E_{\rho}-E_0=\frac{1}{a\sqrt{y}}[\frac{1}{4}+6y-203y^2]\\
\nonumber
&& m_{b_1}=E_{b_1}-E_0=\frac{1}{a\sqrt{y}}[\frac{1}{4}+10y-267y^2]\\
\nonumber
&& m_{a_1}=E_{a_1}-E_0=\frac{1}{a\sqrt{y}}[\frac{1}{4}+14y-1435y^2]\\
\nonumber
&& m_{f_2}=E_{f_2}-E_0=\frac{1}{a\sqrt{y}}[\frac{1}{4}+14y-875y^2]\\
\nonumber &&
m_{f_0}=E_{f_0}-E_0=\frac{1}{a\sqrt{y}}[\frac{1}{4}+18y-1083y^2]
\end{eqnarray}
where $y=1/g^4$.

\subsection{Lattice versus continuum}

The series are valid only for $g^2$ large. To compare the results
of the strong-coupling expansion with the continuum theory we need
to extrapolate these series to the region in which $y>>1$. This
region corresponds to the continuum limit because
$g^4a^4\rightarrow 0$ when $y\rightarrow\infty$. One may consider
the mass ratios, expand them as power series in $y$ and then use
$[1,1]$ Pad\'e approximants by writing the mass ratios in the form
$P_1^1=(1+ay)/(1+by)$ where $a$ and $b$ are determined by
expanding to order $y^2$ and equating coefficients. In the
continuum limit this ratio yields $a/b$. The results obtained with
this method are
\begin{eqnarray}
\nonumber && \frac{m_{\pi_0}}{m_{b_1}}\rightarrow 0.75 \qquad
(\frac{m_{\pi_0}}{m_{b_1}})_{exp}\rightarrow 0.11\\ \nonumber &&
\frac{m_{\omega}}{m_{b_1}}\rightarrow 0.75 \qquad
(\frac{m_{\omega}}{m_{b_1}})_{exp}\rightarrow 0.63\\ \nonumber &&
\frac{m_{\rho}}{m_{b_1}}\rightarrow 0.71 \qquad
(\frac{m_{\rho}}{m_{b_1}})_{exp}\rightarrow 0.62\\ \nonumber
&&\frac{m_{b_1}}{m_{f_0}}\rightarrow 0.82 \qquad
(\frac{m_{b_1}}{m_{f_0}})_{exp}\rightarrow 0.90\\
\nonumber && \frac{m_{b_1}}{m_{a_1}}\rightarrow 0.95 \qquad
(\frac{m_{b_1}}{m_{a_1}})_{exp}\rightarrow 0.98\\ \nonumber &&
\frac{m_{b_1}}{m_{f_2}}\rightarrow 0.92 \qquad
(\frac{m_{b_1}}{m_{f_2}})_{exp}\rightarrow 0.97
\end{eqnarray}
For each mass ratio considered here, the $[1,1]$ Pad\'e
approximant exists with positive values for $a/b$. Therefore, an
extrapolation from $y=0$ to $y\rightarrow\infty$ is singularity
free in this approximation. The results are in very good agreement
with experiment except for the pion mass: this should not surprise
due to the lack of a full chiral symmetry in the lattice theory.


\begin{thebibliography}{9}

\bibitem{'tHooft:1973jz} G.~'t Hooft,
Nucl.\ Phys.\ B {\bf 72}, 461 (1974).

\bibitem{'tHooft:1974hx} G.~'t Hooft,
Nucl.\ Phys.\ B {\bf 75}, 461 (1974).

\bibitem{Teper:2002kh}
M.~Teper,
arXiv:hep-ph/0203203.

\bibitem{Aharony:1999ti} O.~Aharony, S.~S.~Gubser, J.~M.~Maldacena,
H.~Ooguri and Y.~Oz,
Phys.\ Rept.\  {\bf 323}, 183 (2000) [arXiv:hep-th/9905111].

\bibitem{dona}G. Grignani, D.Marmottini and P. Sodano, in preparation.

\bibitem{Banks:1976ia} T.~Banks, S.~Raby, L.~Susskind, J.~B.~Kogut,
D.~R.~Jones, P.~N.~Scharbach and D.~K.~Sinclair [Cornell-Oxford-Tel
Aviv-Yeshiva Collaboration],
Phys.\ Rev.\ D {\bf 15}, 1111 (1977).

\bibitem{Witten:1979kh} E.~Witten,
Nucl.\ Phys.\ B {\bf 160}, 57 (1979).

\bibitem{Berruto:1997jv}
F.~Berruto, G.~Grignani, G.~W.~Semenoff and
P.~Sodano,
Phys.\ Rev.\ D {\bf 57}, 5070 (1998) [arXiv:hep-lat/9710066],
F.~Berruto, G.~Grignani, G.~W.~Semenoff and
P.~Sodano,
Phys.\ Rev.\ D {\bf 59}, 034504 (1999) [arXiv:hep-th/9809006],
F.~Berruto, G.~Grignani, G.~W.~Semenoff and
P.~Sodano,
Annals Phys.\  {\bf 275}, 254 (1999)
[arXiv:hep-th/9901142].

\bibitem{Berruto:1999ir}
F.~Berruto, E.~Coletti, G.~Grignani and P.~Sodano,
Nucl.\ Phys.\ Proc.\ Suppl.\  {\bf 83}, 685 (2000)
[arXiv:hep-lat/9909027],
F.~Berruto, G.~Grignani and P.~Sodano,
Phys.\ Rev.\ D {\bf 62}, 054510 (2000)
[arXiv:hep-lat/9912038].

\bibitem{Kluberg-Stern:1983dg}
H.~Kluberg-Stern, A.~Morel, O.~Napoly and B.~Petersson,
Nucl.\ Phys.\ B {\bf 220}, 447 (1983).

\bibitem{Susskind:1976jm}
L.~Susskind,
Phys.\ Rev.\ D {\bf 16}, 3031 (1977).

\bibitem{Langmann:hd}
E.~Langmann and G.~W.~Semenoff,
Phys.\ Lett.\ B {\bf 297}, 175 (1992).

\end{thebibliography}
\end{document}